\documentclass[a4paper, amsfonts, amssymb, amsmath, reprint, showkeys, nofootinbib, twoside, aps, pre, superscriptaddress]{revtex4-1}

\usepackage{mathptmx}

\usepackage[english]{babel}
\usepackage[utf8]{inputenc}

\usepackage{float}

\usepackage{natbib}
\usepackage[colorinlistoftodos, color=green!40, prependcaption]{todonotes}

\usepackage{indentfirst}

\usepackage[pdftex, pdftitle={Article}, pdfauthor={Author}]{hyperref} 

\usepackage[figurename=FIG., tablename=Table]{caption}

\usepackage{titlesec}
\titlespacing*{\section}{0pt}{5ex}{1ex}

\begin{document}

% date
\date{\today}

% title
\title{Correlated rigidity percolation in fractal lattices}

% authors and affiliations
\author{Shae Machlus}
\affiliation{Department of Physics, University of Chicago, Chicago, Illinois, 60637, USA}
\author{Shang Zhang}
\author{Xiaoming Mao}
\affiliation{Department of Physics, University of Michigan, Ann Arbor, Michigan, 48109, USA}
\captionsetup[figure]{labelsep=period,justification=raggedright}
\captionsetup[table]{labelsep=period,justification=raggedright}

\begin{abstract}
	Rigidity percolation (RP) is the emergence of mechanical stability in networks. Motivated by the experimentally observed fractal nature of materials like colloidal gels and disordered fiber networks, we study RP in a fractal network where intrinsic correlations in particle positions is controlled by the fractal iteration. Specifically, we calculate the critical packing fractions of site-diluted lattices of Sierpi\'nski gaskets (SG’s) with varying degrees of fractal iteration. Our results suggest that although the correlation length exponent and fractal dimension of the RP of these lattices are identical to that of the regular triangular lattice, the critical volume fraction is dramatically lower due to the fractal nature of the network. Furthermore, we develop a simplified model for an SG lattice based on the fragility analysis of a single SG\@. This simplified model provides an upper bound for the critical packing fractions of the full fractal lattice, and this upper bound is strictly obeyed by the disorder averaged RP threshold of the fractal lattices. Our results characterize rigidity in ultra-low-density fractal networks.	
\end{abstract}

\maketitle
% END

% BEGIN INTRODUCTION
\section{INTRODUCTION}
Soft disordered solids are ubiquitous; they exist in many forms such as colloidal gels, fiber networks, colloidal glasses, emulsions, aerogels, polymer melts, and foams. These classes of materials make up biological tissues, food products, cosmetic products, and materials like paper and nonwoven fabric. Some of these soft materials need only a very low density of solid particles to become rigid. In particular, colloidal gels can exhibit nonzero shear rigidity at a wide range of volume fractions \cite{trappe_jamming_2001,trappe_colloidal_2004,Colombo2014,tsurusawa_direct_2019, hsiao_role_2012, wufsus_hydraulic_2013,michel_structural_2019,cho_emergence_2020}, which can be below 1\% in the case of blood clots \cite{wufsus_hydraulic_2013}. 

Classical RP problems are concerned with the emergence of rigidity in discrete mechanical networks. They have been studied in a number of lattices as models of rigidity transitions in soft matter.  
%is a group of discrete lattice models that has been used to capture the emergence of rigidity in soft matter. 
In these models one typically starts with an empty lattice and populates bonds or sites randomly while observing the emergence of a percolating cluster that \emph{can carry stress}.   In comparison with percolation (sometimes called ``geometric percolation''), rigidity percolation not only requires the emergence of an infinite cluster, but also requires that stress can be transmitted from edge to edge of the whole lattice via this infinite cluster.  For example, on a two-dimensional site-diluted triangular lattice, the percolation threshold is $1/2$, and the rigidity percolation threshold is about $69.8\%$ in terms of the fraction of site occupancy \cite{jacobs_generic_1996}.

Classical RP transitions are associated with high values of critical volume fractions $\phi_c$  for a material to be rigid (typically much greater than $10\%$) \cite{jacobs_generic_1996,chubynsky_algorithms_2007,ellenbroek_rigidity_2011,Zhang2015,zhang_correlated_2019}, so how can these ultra-low-density materials exhibit rigidity? Previous work suggested that the answer to this question lies in how the particles are spatially correlated to each other--the Warren truss, for example, transmits stress very efficiently and can achieve rigidity at $\phi_c = 0$ when viewed as a two or three dimensional structure \cite{zhang_correlated_2019}. While colloids will not spontaneously form in Warren trusses (as that involves an unrealistic amount of correlation), moderate correlation strength is still successful in lowering $\phi_c$. While the type of correlation used in \cite{zhang_correlated_2019} was not enough for describing rigidity in ultra-low-density solids, it suggested that there may be another sort of spatial correlation that is both physically realistic and allows the system to achieve an arbitrarily low value of $\phi_c$. We conjecture that a recursive correlation (which generates a fractal network) would be a promising candidate for describing rigidity at ultra-low-densities because (i) fractals are low density while still being connected, and they can be rigid, and (ii) experimental evidence suggests that low density disordered solids (coagulated blood, for example) can indeed be fractal as a result of the non-equilibrium process in which the material is assembled \cite{Carpineti1992,trappe_jamming_2001, gisler_strain_1999, evans_gel_2010, segre_glasslike_2001, vermant2005flow,cho_emergence_2020}.  

In this paper, we show that a model fractal network, the Sierpi\'nski gasket lattice (SGL), does indeed achieve rigidity at arbitrarily low volume fractions. The SGL exhibits intrinsic positional correlation between the particles which increases with its number of fractal iterations $n$. This result is supported analytically by simple calculation on the undiluted SGL and numerically on the randomly diluted SGL by using the pebble game algorithm. We also calculate the correlation length and fractal dimension critical exponents for RP in this lattice and find that the universality class of the rigidity phase transition in the lattice is the same as that for the regular triangular lattice. We further propose a simple non-fractal model, the RP of which yields a strict upper bound to the disorder-averaged critical volume fraction of the SGL.
% END

% BEGIN MODEL
\section{MODEL} 
\begin{figure}
    \includegraphics[width=8cm]{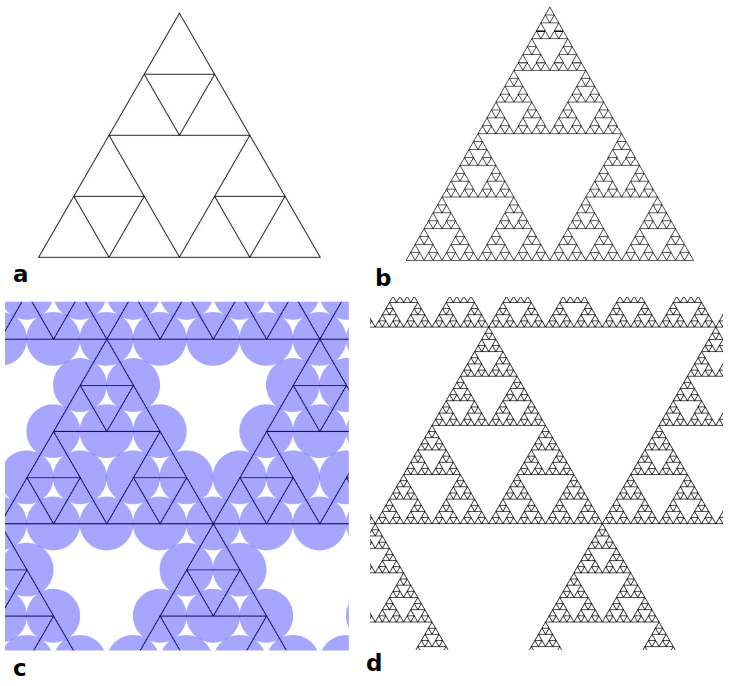}
	\caption{(a,b) Sierpi\'nski gasket (SG) of fractal iteration $n=2,5$. (c,d) Lattices of SG's are models for ultra-low-density networks at $n=2,5$. In (c) semi-transparent purple disks represent the physical particles we are modeling. The diameter of each particle is equal to the bond length, which we set to 1.}
	\label{FIG:SGgrid}
\end{figure}

We use a lattice that achieves an arbitrarily low volume fraction while still exhibiting rigidity at full site occupancy. Motivated by the experimentally observed fractal structure of fiber networks and colloidal gels \cite{witten_diffusion-limited_1981, gisler_strain_1999, cho_emergence_2020}, we consider a triangular lattice where the upwards pointing triangles are replaced by Sierpi\'nski gaskets (SG's), as shown in Fig.~\ref{FIG:SGgrid}. The unit cell of this lattice is an upwards pointing SG with an adjoining vacant downwards pointing triangle, which forms a rhombus. Vibrational modes and spin phase transitions have been studied on this lattice \cite{gefen_phase_1984, luscombe_statistical_1985, da_rocha_transition_2011, yu_vibrational_1984, burioni_vibrational_2002,liu_spectral_1985,liu_spectral_1984}. This is a rich lattice to study since there are three length scales: (i) the size of the smallest triangle in an SG which we always set as $1$, (ii) the length of the edge of an SG $2^n$, and (iii) the length of the lattice $L=s2^n$. $s$ is the number of SG's on one side of the lattice, and $n$ is the number of times the SG pattern repeats on itself, what we call the fractal iteration number. We emphasize that $L$ is measured in units of the smallest triangle of an SG since the length of the smallest triangle is always 1, independent of $n$. Also note that $n=0$ corresponds to a regular triangular lattice. 

The volume fraction of the SGL, at full site occupancy, is 
\begin{equation}
\phi_{\textnormal{SGL}_{\textnormal{undiluted}}} = A \frac{3^{n+1}-1}{2^{2n}}, 
\end{equation}
where the constant $A$ is the area of the particle. In the SGL we consider here, $A=\pi/4\sqrt{3}$.
This result is derived in Appendix~\ref{APP:calculating_phi}, and it is obtained by assuming that each site is occupied by a disk whose diameter equals the bond length between neighboring sites, pictured in Fig.~\ref{FIG:SGgrid}(c). It follows that 
\begin{equation}
\lim_{n \rightarrow \infty} \phi_{\textnormal{SGL}_{\textnormal{undiluted}}} (n) = 0 .
\end{equation}
An arbitrarily large $n$ corresponds to an arbitrarily small $\phi_{\textnormal{SGL}_\textnormal{undiluted}}$, so the SGL is indeed a suitable model to study the emergence of rigidity in ultra-low-density networks. A single SG, of any $n$, is isostatic--it has 3 trivial zero modes and no states of self stress \cite{lubensky_phonons_2015}. The coordination number of the undiluted lattice under periodic boundary conditions $\langle z \rangle_{\textnormal{undiluted}}$ can be calculated as a function of $n$. 
\begin{equation}
\langle z \rangle_{\textnormal{undiluted}} = \frac{6 + 4(x-1)}{x} ,
\label{EQ:z}
\end{equation}
where $x=(3^{n+1}-1)/2$ is the number of sites present in a single $n$-level SG where $n \geq 1$. At $n=0$, the lattice is a regular triangular lattice, so $\langle z \rangle_{\textnormal{undiluted}} = 6$. The coordination number decreases from 6 to 4 as $n$ goes from 0 to $\infty$. 

We dilute the SGL by removing randomly chosen sites. If a site is removed, all of the bonds attached to that site are also removed. The occupancy fraction $p_{\textnormal{SGL}}$ is the ratio of the number of occupied sites to the number of sites present in a completely filled SGL. As shown in Appendix~\ref{APP:calculating_phi}, the volume fraction of the diluted SGL is then
\begin{equation}\label{EQ:phip}
\phi_{\textnormal{SGL}} = p_{\textnormal{SGL}} \phi_{\textnormal{SGL}_{\textnormal{undiluted}}} .
\end{equation}

We emphasize that while the occupancy fraction $p_{\textnormal{SGL}}$ is the ratio of the number of occupied sites to total number of sites (unoccupied and occupied), the volume fraction $\phi_{\textnormal{SGL}}$ is the ratio of the occupied space to the total space covered by the lattice. Because the volume fraction of the undiluted SGL $\phi_{\textnormal{SGL}_{\textnormal{undiluted}}}$ vanishes in the $n \to \infty$ limit, $\phi_{\textnormal{SGL}}$ can approach $0$ even when $p_{\textnormal{SGL}}$ is of $\mathcal{O}(1)$.

% END

% BEGIN METHOD AND RESULTS
\section{METHOD \& RESULTS}
A rigid cluster in a mechanical network is a collection of sites and bonds without any floppy modes. The only zero energy normal modes of a rigid cluster are trivial rigid-body degrees of freedom of the whole cluster.   If a rigid cluster spans the whole network, the system must exhibit at least one positive elastic modulus. The emergence of such an infinite rigid cluster is called RP. It is worth mentioning that when rigidity percolates, floppy modes may still exist in other parts of the network which are not in an infinite rigid cluster. The pebble game is an efficient algorithm that can be used to examine rigidity in two dimensions~\cite{jacobs_generic_1995, jacobs_algorithm_1997}.  
%algorithm determines whether a cluster is rigid or not by placing a ``test" bond between two sites in the cluster. 
%If there are exactly four degrees of freedom in the cluster which can be localized to the test bond, then the inclusion of that bond makes the cluster minimally rigid, or isostatic. If there are less than four degrees of freedom available, the cluster is already rigid, and including the test bond is redundant to the rigidity of the cluster. If there are more than four degrees of freedom which may be localized at that bond, then the cluster is not presently rigid, nor does it become rigid upon the inclusion of the test bond. As more sites, and thus more bonds, are introduced to the system, the size of the largest rigid cluster grows until it spans the system.

In order to study the RP in the diluted SGL, we execute the pebble game algorithm  on SGL's at $n=1,2,3,4,5$ with periodic boundary conditions. For each value of $n$, we consider 4 different system sizes $L$ which were chosen so that the lattices have approximately 250, 1000, 4000, and 16,000 particles (sites) (although at $n=5$ we consider only the 3 larger system sizes because each SG at $n=5$ already contains a large number of sites, and we need to keep the number of SG's large in the lattice). To keep the number of sites roughly constant across varying $n$, we reference
\begin{equation}
 L = 2^{\frac{2n+1}{2}} \sqrt{\frac{N}{3^{n+1}-1}},
\end{equation}
which is immediate from Eqs. (\ref{EQ:sites}) and (\ref{EQ:length}) (Appendix~\ref{APP:calculating_phi}), to choose an integer valued side length $L$ for each target system size (in terms of the total number of sites) and fractal iteration $n$.
%We say this equation is ``referenced'' instead of ``used'' because the right hand side of the equation must be adjusted to a multiple of the side length of an SG of fractal iteration $n$ so that each lattice has an integer number of SG's. 

For each $n$ and $L$, we generate 200 samples of SGL's. Each one represents a realization of disordered dilution. For each SGL, initially empty, we add new sites randomly to the lattice one by one. Each new site added increases $p_{\textnormal{SGL}}$. We run the pebble game algorithm at regular intervals of $p_{\textnormal{SGL}}$ on this lattice to determine when a spanning rigid cluster appears. The occupancy fraction at which this occurs is the critical occupancy fraction $p_{c,\textnormal{SGL}}$. We record the mass of the spanning rigid cluster $M_{c,\textnormal{SGL}}$ when it first occurs in each sample. The code used to produce this data is contained in a GitHub repository~\cite{github_link}.
%As a percolation problem, $M_{c,\textnormal{SGL}}$ is the order parameter of the transition.  
We then average over the 200 samples to obtain the averaged quantities, $\langle M_{c,\textnormal{SGL}}(n,L)\rangle$ and $\langle p_{c,\textnormal{SGL}}(n,L) \rangle$, for each $n$ and $L$.  We also measure the fluctuation of the transition point
%As a percolation problem, the mass of the spanning rigid cluster is the order parameter; the cluster grows and spreads as more sites are occupied. For each $n$ and $L$ we average all the $M_{c,\textnormal{SGL}}$'s and $p_{c,\textnormal{SGL}}$'s over these 200 samples to obtain $\langle M_{c,\textnormal{SGL}}(n,L)\rangle$ and $\langle p_{c,\textnormal{SGL}}(n,L) \rangle$. We also measure the fluctuation of the transition point
\begin{equation}
\Delta_{p_{c,\textnormal{SGL}}} = \sqrt{\langle p_{c,\textnormal{SGL}}(n,L)^2 \rangle - \langle p_{c,\textnormal{SGL}}(n,L) \rangle ^2} .
\end{equation}
%Using the averaged quantities, 
Our previous study of correlated RP on the triangular lattice~\cite{zhang_correlated_2019} showed that the short-range spatial correlation only shifts the transition point and does not change the universality class of RP in the triangular lattice.  Following this result, we make the assumption that RP in the SGL is also a continuous transition, with the mass of the infinite rigid cluster being the order parameter. This assumption is verified by our scaling results below. 

We invoke finite-size scaling relations \cite{stauffer_introduction_1994, zhang_correlated_2019} to calculate the critical exponents associated with the rigidity phase transition. The correlation length exponent $\nu_{\textnormal{SGL}}$ and the fractal dimension $d_{f,\textnormal{SGL}}$ are calculated as the slopes of linear fits of log-log plots of $\langle M_{c,\textnormal{SGL}} \rangle$ and $\Delta_{p_{c},\textnormal{SGL}}$ versus $L$, according to the finite size scaling relations
\begin{equation}\label{EQ:df}
\langle M_{c,\textnormal{SGL}}(n,L) \rangle \propto L^{d_{f,\textnormal{SGL}}} ,
\end{equation}
\begin{equation}\label{EQ:nu}
\Delta_{p_c,\textnormal{SGL}} \propto L^{-1/\nu_{\textnormal{SGL}}} 
\end{equation}
(Appendix \ref{APP:calc_crit_exp}).
Note that these relations give a calculation of $d_{f,\textnormal{SGL}}$ and $\nu_{\textnormal{SGL}}$ for each $n$.

\begin{figure}
    \includegraphics[width=7cm]{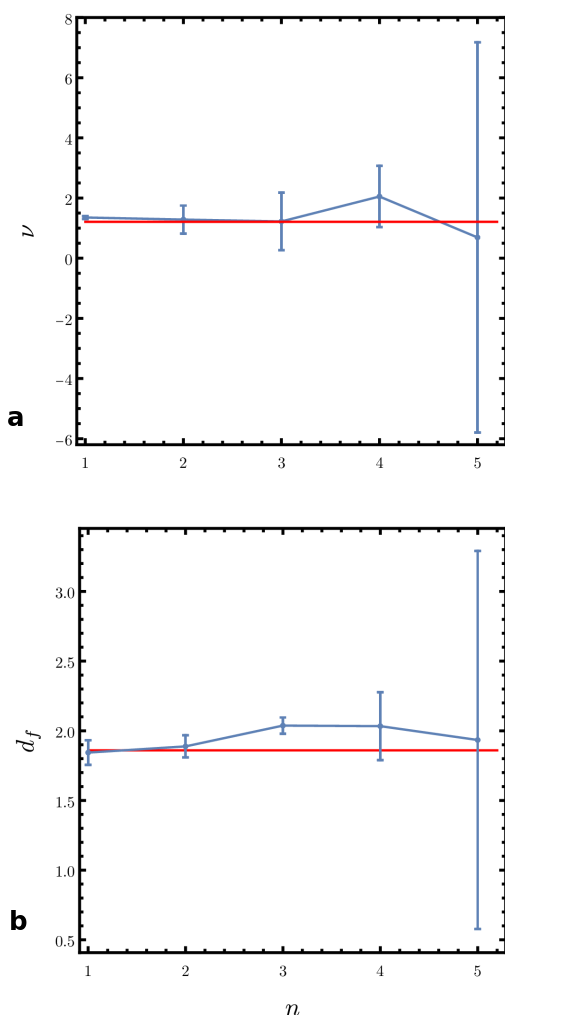}
	\caption{(a) The correlation length exponent $\nu_{\textnormal{SGL}}$ at $n=1,2,3,4,5$. (b) The spanning rigid cluster fractal dimension $d_{f,\textnormal{SGL}}$ for the five values of $n$. The red lines show these exponents for classical RP in the regular triangular lattice ($\nu = 1.21 \pm 0.06$ and $d_f = 1.86 \pm 0.02$) \cite{jacobs_generic_1996}. The error bars are 95\% confidence intervals.}
	\label{FIG:exp_graphs}
\end{figure}

We find $\nu_{\textnormal{SGL}}$ and $d_{f,\textnormal{SGL}}$ for the SGL rigidity phase transition are the same as for the rigidity phase transition in the regular triangular lattice \cite{jacobs_generic_1996} as shown in Fig.~\ref{FIG:exp_graphs}. This observation is consistent with results on RP in lattices with spatial correlations \cite{zhang_correlated_2019}, where the critical exponents remain the same as in classical RP, and the short-ranged spatial correlation can be viewed as an irrelevant perturbation. Here, the fractals in each unit cell can also be viewed as a short range feature, which do not change the divergent length scale at the transition. 
We assert that the large-scale fractal structure of the spanning rigid cluster in the infinite system size limit overwhelms the local fractal structure of the SG's, so $d_{f,\textnormal{SGL}}$ is the same as in the regular triangular lattice case instead of being the fractal dimension of the SG. We also verify that our assumption (the phase transition is continuous) is well justified since the phase transition belongs to the same universality class as \cite{zhang_correlated_2019}. 

We extract the critical occupancy fraction at the infinite system size limit $p_{c,\textnormal{SGL}}(n,L=\infty)$ by linearly extrapolating the finite critical occupancy fractions $p_{c,\textnormal{SGL}}(n,L)$ for each $n$ as a function of $L^{-1/\nu}$. The $p_{c,\textnormal{SGL}}(n,L=\infty)$ are simply the y-intercepts of these linear fits which are displayed in Fig.~\ref{FIG:infinite_size}. Further information about this process can be found in Appendix C of \cite{zhang_correlated_2019}. 

\begin{figure}
	\includegraphics[width=8cm]{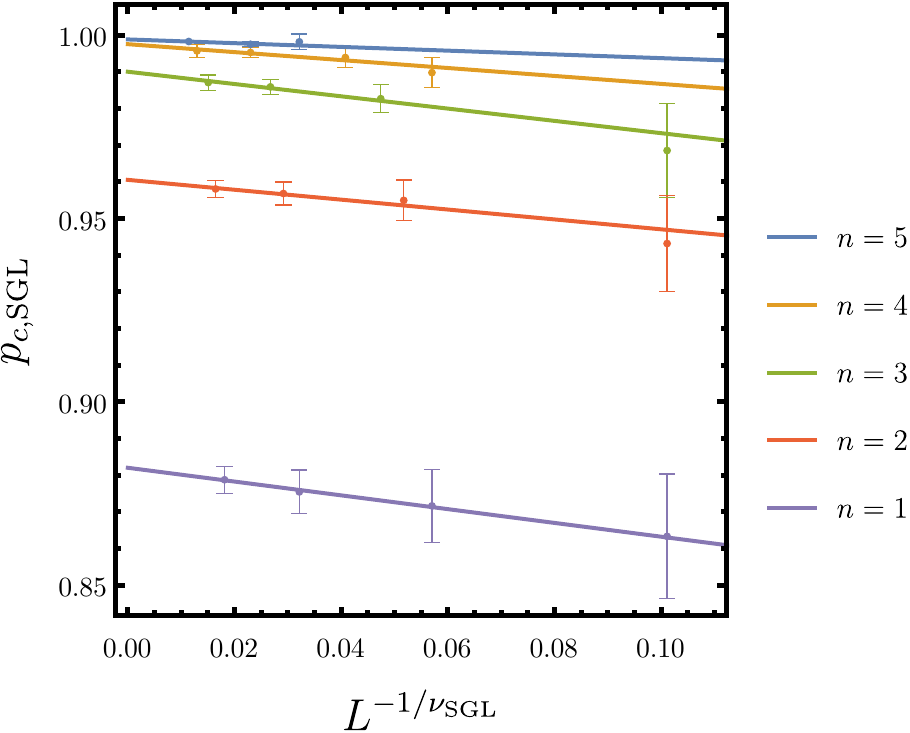}
	\caption{Extracting $p_{c,\textnormal{SGL}}(n,L=\infty)$ from the linear extrapolation of the finite-size critical occupancy fractions $p_{c,\textnormal{SGL}}(n,L)$ as a function of $L^{-1/\nu_{\textnormal{SGL}}}$ where $\nu_{\textnormal{SGL}} = 1.21$. The lines are linear fits, and the y-intercepts are the infinite-size limit of the critical occupancy fractions $p_{c,\textnormal{SGL}} (n, L = \infty)$. The error bars are 95\% confidence intervals.}
	\label{FIG:infinite_size}
\end{figure}
We find that the critical occupancy fraction $p_{c,\textnormal{SGL}}(n,L=\infty)$ approaches 1 as $n$ increases while the critical volume fraction $\phi_{c,\textnormal{SGL}}(n,L=\infty)$ approaches 0 [following the relation in Eq.~\eqref{EQ:phip}], indicating that these disordered fractal structures exhibit rigidity at vanishing volume fractions. These results are shown in Table~\ref{TAB:p_and_phi}.
\begin{table}
	\begin{tabular}{|c|c|c|}
	\hline
	$n$ & $p_{c,\textnormal{SGL}}(n,L=\infty)$ & $\phi_{c,\textnormal{SGL}}(n,L=\infty)$ \\
	\hline
	1 & $0.882 \pm 0.002$ & $0.800 \pm 0.002$ \\
	2 & $0.961 \pm 0.004$ & $0.708 \pm 0.003$ \\
	3 & $0.990 \pm 0.004$ & $0.561 \pm 0.002$ \\ 
	4 & $0.998 \pm 0.004$ & $0.428 \pm 0.002$ \\
	5 & $0.999 \pm 0.005$ & $0.322 \pm 0.002$ \\
	\hline	
	\end{tabular}
	\caption{The critical occupancy and volume fractions for the SGL's for $n=1,2,3,4,5$ in the infinite system size limit, $p_{c,\textnormal{SGL}}(n,L=\infty)$ and $\phi_{c,\textnormal{SGL}}(n,L=\infty)$. As $n$ increases, $p_{c,\textnormal{SGL}}(n,L=\infty) \rightarrow 1$ and $\phi_{c,\textnormal{SGL}}(n,L=\infty)\rightarrow 0$. The error values are 95\% confidence intervals.}
	\label{TAB:p_and_phi}
\end{table}
% END

% BEGIN INTERPRETATION
\section{INTERPRETATION} The fact that the $p_{c,\textnormal{SGL}}(n,L=\infty)$'s approach 1 as $n$ increases is a reflection of both the fragility of a single SG--for any value of $n$, removing any non-corner site of an SG segregates the three corners of the SG into three separate rigid clusters (Appendix~\ref{APP:frag}), and the result [Eq.~\ref{EQ:z}] that $\langle z \rangle$ approaches the critical value of 4 as $n$ increases. The latter point reveals that the SGL is asymptotically a Maxwell lattice (i.e., lattices that satisfy $\langle z \rangle=2d$ and are thus at the verge of mechanical instability~\cite{mao_maxwell_2018,lubensky_phonons_2015}) as $n \rightarrow \infty$.

\begin{figure}
	\includegraphics[width=6cm]{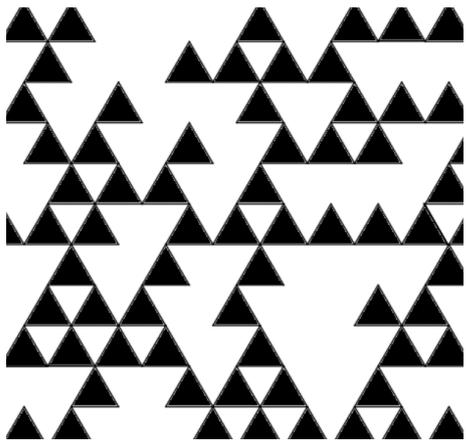}
	\caption{The triangular plate model (TPL) is a regular triangular lattice which has been diluted in units of upwards pointing equilateral triangles (black).}
	\label{FIG:TPL}
\end{figure}

These observations motivate a simplified model of the SGL--the triangle plate lattice (TPL). The TPL is a regular triangular lattice consisting of upwards-pointing rigid triangles hinged at their tips. In other words, if we view it as a regular bond-dilution RP in a triangular lattice, the items which are being diluted are groups of three bonds which together form an upwards pointing triangle. Figure~\ref{FIG:TPL} is an example of what a diluted TPL can look like. 

There is one main feature that separates the TPL from the SGL: in the SGL an SG with a site removed may still be an essential part of the spanning rigid cluster. In the TPL, a vacant triangle cannot transmit rigidity. Because of this difference the critical packing fraction of the TPL is used to calculate \textit{a strict upper bound} on that of the SGL.

All $p$'s that follow in this section should be taken to be in the infinite system size limit. The relationship between $p_{c,\textnormal{SGL}}$ and the critical packing fraction for the TPL $p_{c,\textnormal{TPL}}$ is as follows: consider an SGL and a TPL, where the SG's in the SGL and the triangle plates in the TPL are the same size. Let the two lattices also be of equal size. A removed upwards pointing triangle from the TPL corresponds to \textit{at least} one removed site from the SGL. Letting the number of triangles/SG's present in either lattice be $N_{\Delta}$ and the number of sites present in a single SG be $x=(3^{n+1}-1)/2$, the critical occupancy fractions for the two lattices are related by
\begin{equation}
xN_{\Delta}(1-p_{c,\textnormal{SGL}}) \geq N_{\Delta}(1-p_{c,\textnormal{TPL}}) .
\end{equation}
The number of removed sites at the critical point in the SGL is at least the number of removed triangles at the critical point in the TPL\@. The ``=" sign is only satisfied if removing each site from the SGL corresponds to removing a distinct triangle plate from the TPL. This is not always the case because (i) multiple removed sites in the SGL can belong to the same SG, and, as we discussed above, (ii) a ``broken" SG can still contribute to the rigidity of the lattice.  As a result, the TPL provides an upper bound of the critical occupancy in the SGL, $P_{c,\textnormal{SGL}}$. Explicitly,
\begin{equation}
p_{c,\textnormal{SGL}} \leq 1 - \frac{1-p_{c,\textnormal{TPL}}}{x} \equiv P_{c,\textnormal{SGL}}.
\end{equation}

\begin{figure}
	\includegraphics[width=7cm]{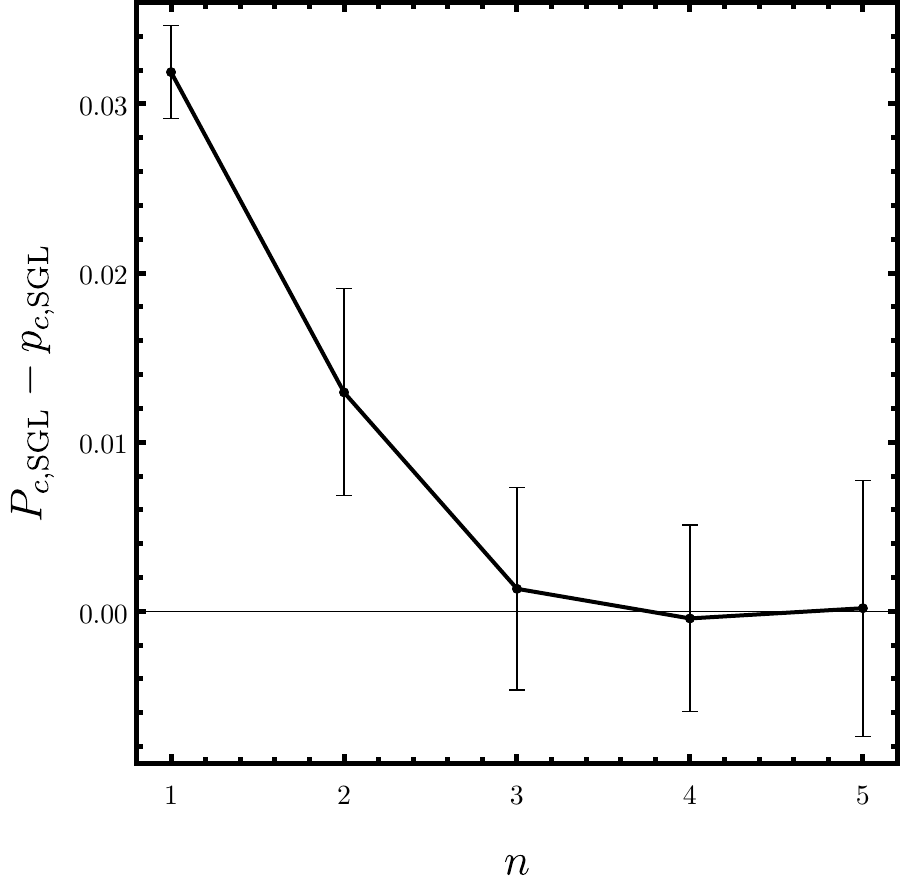}
	\caption{The difference between the upper bound on $p_{c,\textnormal{SGL}}$ given by the TPL, $P_{c,\textnormal{SGL}}$, and the measured $p_{c,\textnormal{SGL}}$ becomes smaller as $n$ increases. The error bars are 95\% confidence intervals.}
	\label{FIG:model_results}
\end{figure}

We perform the pebble game routine on the TPL and execute the same finite scaling procedures that we did for the SGL. We find that $p_{c,\textnormal{TPL}} = 0.656 \pm 0.005$ and $\nu_{\textnormal{TPL}} = 1.4 \pm 0.1$. The errors given are 95\% confidence intervals. $p_{c,\textnormal{TPL}}$ and $\nu_{\textnormal{TPL}}$ both lie within error bars of the corresponding variables for the regular triangular lattice in the case of bond dilution \cite{jacobs_generic_1996}. The upper bounds on the $p_{c,\textnormal{SGL}}$'s predicted by the TPL are obeyed for all tested values of $n$ and tightly obeyed for larger values of $n$ (Fig.~\ref{FIG:model_results}). It is worth pointing out that this is a strict upper bound in the sense of disorder averaged critical occupancy. It does not necessarily hold for individual samples.
%END

% BEGIN CONCLUSIONS AND DISCUSSIONS
\section{CONCLUSION, DISCUSSIONS, \& EXTENSIONS}
In this paper we show that by introducing fractal local structures, rigidity can exist at an arbitrarily low volume fraction of solid particles. Using a periodic lattice model consisting of Sierpi\'nski gaskets, we find that as the fractal iteration increases, the critical site occupancy fraction for rigidity increases, while the critical volume fraction decreases, allowing rigidity at progressively lower volume fractions. We also show that the RP transition in this fractal lattice remains in the same universality class as the classical RP transition when length is measured in units of the sides of the smallest triangles. We interpret this result by mapping the RP on this fractal lattice into the RP of a simple triangle plate model, based on the fragility of a single SG.  This mapping gives a strict upper bound of the critical volume fraction of the fractal lattice.

Our results may shed light on the origin of rigidity in ultra-low volume fraction soft solids, such as hydrogels and aerogels. A simple way to understand this phenomena is to realize that, even in a dense disordered solid such as granular matter or colloidal glass, stress is often carried by a very small fraction of the solid content, i.e.,  \emph{force chains}~\cite{cates1998jamming,bi2011jamming,Zhang2017}, while other components do not significantly contribute to the elasticity.  Thus, by introducing appropriate spatial correlation between the solid particles, a material can be constructed without filling the space which is not needed for rigidity.  Interestingly, interactions and non-equilibrium processes (such as hydrodynamics of the solvent) occuring during the formation of these ultra-low volume fraction solids appear to naturally achieve this goal of arranging particles in very efficient ways of transmitting stress.  It is of our interest to understand how this occurs in these experimental systems in the future.

The model we discuss here is a two-dimensional lattice. A curious question that immediately arises is what happens in three dimensions. The SG has a direct three-dimensional generalization: the Sierpi\'nski tetrahedron (ST), which is constructed by iteratively hinging tips of four tetrahedra together to form a bigger tetrahedra (which has an octahedron of empty space in the middle). Each face of an ST is an SG. Interestingly, there is a mechanical analogy between the SG and the ST: each internal node in the ST has six bonds, satisfying the Maxwell condition $\langle z \rangle =2d$, while the four tip nodes each have three bonds ($z=3$), giving rise to exactly the six trivial rigid body motions of the whole ST. Thus, each ST is isostatic in three dimensions.  

These ST's can be used to construct a face-centered-cubic lattice in the same way the SG's are used to construct the SGL. This three-dimensional lattice also has a volume fraction that approaches zero as its fractal iteration increases. Analogously, in the undiluted face-centered-cubic lattice, each node at the tip of an ST has $z=12$, taking the whole structure to $\langle z \rangle >6$. It is straightforward to see that the undiluted ST lattice has rigidity from the rigidity of the single ST's and from the stress-bearing structures (states of self-stress along straight lines of bonds)~\cite{lubensky_phonons_2015,mao_maxwell_2018,Zhang_2018}. Therefore, a similar RP problem can be formulated for this three-dimensional ST lattice. The nature of the RP transition may be more complicated because it is a three-dimensional problem \cite{chubynsky_algorithms_2007}, but this lattice at least provides an example of a three-dimensional lattice where rigidity exists at an arbitrarily low volume fraction. It is also of our interest to study the RP transition in this three-dimensional lattice in the future.
% END

% BEGIN ACKNOWLEDGEMENTS
\section*{ACKNOWLEDGEMENTS}
We thank the National Science Foundation for their support through Grant No. DMR-1609051 (S. Z. and X. M.) and NSF PHY 1852239 ``Summer Undergraduate Research in Physics and Astrophysics at the University of Michigan” (S. M.)
%NSF-EFRI-1741618
% END
\appendix

\begin{figure}[t]
	\includegraphics[width=7cm]{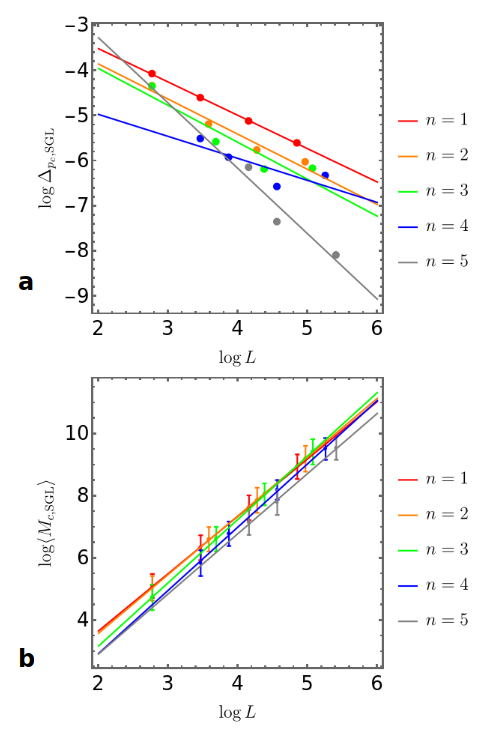}
	\caption{(a) The correlation length exponent $\nu_{\textnormal{SGL}}$ and (b) the fractal dimension $d_{f,\textnormal{SGL}}$ for the SGL are both obtained from the slopes of the linear fits for each $n$ according to Eqs.~\eqref{EQ:df} and~\eqref{EQ:nu}. The error bars are 95\% confidence intervals.}\label{FIG:scaling}
\end{figure}

% BEGIN APPENDIX A
\section{CALCULATING \texorpdfstring{$\phi_{\textnormal{SGL}}$}{phiSGL}}
\label{APP:calculating_phi}
% make the equation numberings appropriate for an appendix according to APS
\renewcommand{\theequation}{A\arabic{equation}}
The volume fraction, an area fraction for $d=2$, is the ratio of space taken up by the occupied sites to the space enclosed within the unit cell. $\phi_{\textnormal{SGL}}$ is the volume fraction of the lattice, $N_{\textnormal{occ}}$ is the number of occupied sites in the lattice, $A_v$ is the area covered by a single site, and $A_l$ is the total area covered by the lattice. 
\begin{equation}
	\phi_{\textnormal{SGL}} \equiv \frac{N_{\textnormal{occ}}A_v}{A_l}.
	\label{phi_def}
\end{equation}
The lattice is a rhombus with side length $L$, so
\begin{equation}
	A_l = \frac{\sqrt{3}}{2}L^2.
\end{equation}
Additionally, we define the occupancy fraction $p_{\textnormal{SGL}}$ as
\begin{equation}
	p_{\textnormal{SGL}} \equiv \frac{N_{\textnormal{occ}}}{N_{\textnormal{total}}},
\end{equation}
where $N_{\textnormal{total}}$ is the total number of sites (occupied and unoccupied) in the lattice. For an SGL with periodic boundary conditions, $N_{\textnormal{total}}$ is given by
\begin{equation}
	N_{\textnormal{total}} = s^2 \left(\frac{3^{n+1} - 1}{2}\right),
	\label{EQ:sites}
\end{equation}
where $n$ is the number of fractal iterations, and $s$ is the length of the lattice in units of SG's. We set the distance between neighboring sites on the lattice to be 1. Due to the fractal structure of an SG,
\begin{equation}
    L = s 2^n.
    \label{EQ:length}
\end{equation}
Since the length between sites is 1, we also know that
\begin{equation}
	A_v = \pi\left(\frac{1}{2}\right)^2.
\end{equation}
Putting everything together,
\begin{equation}
	\phi_{\textnormal{SGL}} = A \frac{3^{n+1}-1}{2^{2n}}p_{\textnormal{SGL}}.
\end{equation}
where the constant $A = \pi/4\sqrt{3}$ is specific to the geometry of the system.
% END

% BEGIN APPENDIX B
\section{CALCULATING CRITICAL EXPONENTS}
\label{APP:calc_crit_exp}
Given the finite size scaling relations Eqs.~\eqref{EQ:df} and~\eqref{EQ:nu}, we can calculate the correlation length exponent $\nu_{\textnormal{SGL}}$ and the fractal dimension $d_{f,\textnormal{SGL}}$ for the SGL, as shown in Fig.~\ref{FIG:scaling}.

% END

% BEGIN APPENDIX C
\section{FRAGILITY OF AN SG}
\label{APP:frag}
We use induction to prove that removing any non-corner site in an SG will segregate the 3 corner sites into different rigid clusters. If a corner site is removed in a free SG, the rigidity of the SG is unaffected. If a corner site is removed in an SGL, the SG's are disconnected, and may not be rigid with respect to one another. 

\begin{figure}
	\includegraphics[width=8.5cm]{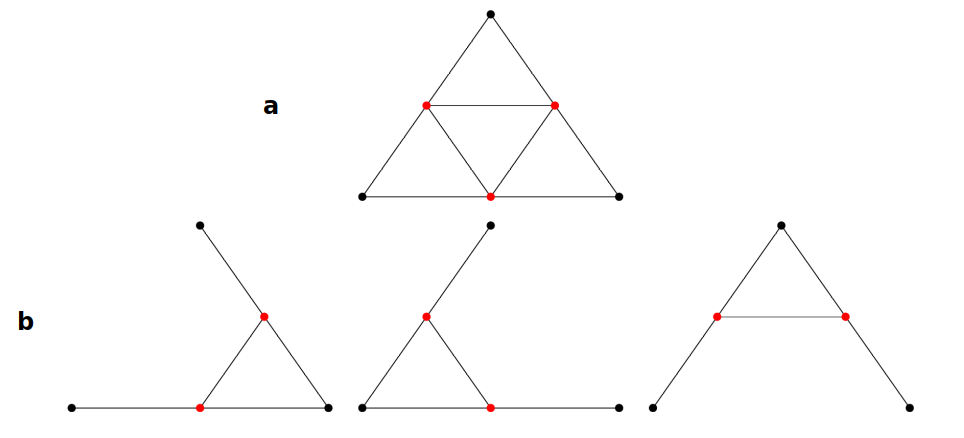}
	\caption{(a) An $n=1$ SG. (b) Removing any non-corner site (red) from an $n=1$ SG leaves two rotors attached to a rigid triangle. The triangle and both rotors (particles with only one bond) each have a corner site (black), so all three corner sites belong to distinct rigid clusters.} \label{FIG:SG1}
\end{figure}

\begin{figure}
	\includegraphics[width=4cm]{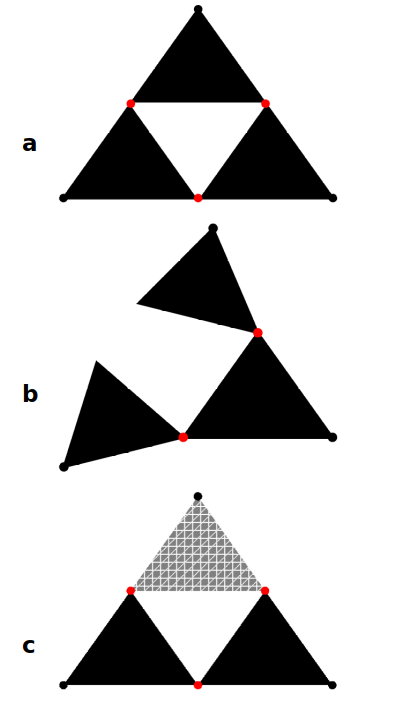}
	\caption{(a) An $(n+1)$-level SG, composed of three $n$-level SG's (black). (b) Case (i), a site connecting two $n$ level $SG$'s (red) is removed, allowing independent motion of the three corner sites (black). (c) Case (ii), a non-coner site is removed from an $n$-level SG (gray with white hatching). If the three corners of the $n$-level SG are in separate rigid clusters, the three corners of the $(n+1)$-level SG can move independently and are thus also in separate rigid clusters.} 
	\label{FIG:SGnplus1}
\end{figure}

Consider an $n=1$ SG. It is immediate from Fig.~\ref{FIG:SG1} that our desired result holds in this case. Suppose this result holds for an $n$-level SG.
Consider now an SG of fractal iteration $n+1$, displayed in Fig.~\ref{FIG:SGnplus1}(a). It is composed of 3 SG's each of fractal iteration $n$. When any internal site of the $(n+1)$-level SG is removed, there are two possible cases: (i) the site is a shared corner site between two $n$-level SG's, shown in Fig.~\ref{FIG:SGnplus1}(b), or (ii) the site is a non-corner site which belongs to a single $n$-level SG, shown in Fig.~\ref{FIG:SGnplus1}(c).
If (i), the two $n$-level SG's which were previously connected are now free to rotate about the hinges they each share with the third unaltered $n$-level SG. The 3 corners of the $(n+1)$-level SG are now in separate rigid clusters. If (ii), then the 3 corners of the $n$-level SG from which a site was removed are now in different rigid clusters, so they can move freely relative to each other. Since the two unaltered SG's are independently rigid, the node connecting the two unaltered SG's is a free hinge, so the three corners of the $(n+1)$-level SG must be in separate rigid clusters. Since assuming our claim is true for an $n$-level SG implies our claim is true for an $(n+1)$-level SG, and the $n=1$ case is manifestly true, for an SG of an arbitrary number of fractal iterations, removing any non-corner site will segregate the 3 corner sites of that SG into different rigid clusters. An SG is ``fragile" in the sense that it has this property.
% END

% BEGIN POSTAMBLE
\bibliography{fractalRP.bib}
\end{document}